\documentclass[a4paper,tbtags,reqno,twoside,12pt]{article}
\usepackage{amsmath}
\usepackage{amsthm}
\usepackage{amssymb}
\newcommand\DN{\newcommand}

\theoremstyle{definition}

\newtheorem{thm}{Theorem}

\newtheorem{rem}{Remark} [section]
\DN\lref[1]{Lemma~\ref{#1}}
\DN\tref[1]{Theorem~\ref{#1}}
\DN\pref[1]{Proposition~\ref{#1}}
\DN\sref[1]{Section~\ref{#1}}
\DN\dref[1]{Definition~\ref{#1}}
\DN\rref[1]{Remark~\ref{#1}} 
\DN\corref[1]{Corollary~\ref{#1}}
\DN\eref[1]{Example~\ref{#1}}
\newcounter{Const} 
\DN\Ct{\refstepcounter{Const}c_{\theConst}}
\DN\cref[1]{c_{\ref{#1}}}

\begin{document}

\begin{center}{\em{\textbf{
Dynamical rigidity of stochastic Coulomb systems \\
in infinite-dimensions}}}\\
\bigskip
{\textrm{
Hirofumi Osada (Kyushu University) }
 }\\
 \medskip
 
 {\small (To appear in RIMS $\mathrm{K\hat{o}ky\hat{u}roku}$)}
 \end{center}

\medskip

\begin{flushright}
{\small 2013/12/18/Wed  \\
at Research Institute of Mathematical Sciences (Kyoto University)} \\
\bigskip

{\small Faculty of Mathematics, Kyushu University \\
Fukuoka, 819-0395, JAPAN \\email:osada@math.kyushu-u.ac.jp }
\end{flushright}

\bigskip
\bigskip

This paper is based on the talk in \lq\lq Probability Symposium" 
at Research Institute of Mathematical Sciences (Kyoto University), and gives an announcement of some parts of the results in \cite{o.sub,o-t.airy,o-t.tail,o-h}.

We consider an infinite-dimensional stochastic dynamics 
$ \mathbf{X}=(X^i)_{i\in\mathbb{N}}$ describing 
infinite-many Brownian particles moving in $ \mathbb{R}^d$ 
interacting through $ \gamma $-dimensional Coulomb potentials $ \Psi _{\gamma}$ with inverse temperature $ \beta $. Here in our definition 
\begin{align}&\label{:0} \quad \quad 
\nabla \Psi_{\gamma } (x) = - \frac{x}{|x|^{\gamma }} 
\quad \quad (x\in\mathbb{R}^d)
.\end{align}
Thus $ \Psi_{\gamma } $ is a special case of Riesz potentials. We will later give a generalization of $ \Psi _{\gamma }$ for $ \gamma \in \mathbb{R}^+\backslash \mathbb{N}$ in \eqref{:P2}; 
we will take $ \Psi _{\gamma }$ as a Riesz potential with 
$ d \le \gamma \le d + 2 $, which is excluded in the classical theory of Gibbs measures based on DLR equations. 

If the stochastic dynamics $ \mathbf{X} $ 
is translation invariant, then $ \mathbf{X}=\{ (X_t^i)_{i\in\mathbb{N}} \}_{t\in[0,\infty)} $ is given by the solution of the infinite-dimensional stochastic differential equation (ISDE): 
\begin{align}\label{:1}&
dX_t^i = dB_t^i + \frac{\beta }{2} \lim_{r\to\infty }
\sum_{j\not=i,\, | X_t^i-X_t^j |<r } 
\frac{ X_t^i-X_t^j }{ | X_t^i-X_t^j |^{\gamma } }dt \quad (i\in\mathbb{N})
,\end{align}
and  equivalently 
\begin{align}\label{:1}&
dX_t^i = dB_t^i - \frac{\beta }{2} \lim_{r\to\infty }
\sum_{j\not=i,\, | X_t^i-X_t^j |<r } 
\nabla \Psi _{\gamma } (X_t^i-X_t^j) dt \quad (i\in\mathbb{N})
.\end{align}
We consider ISDEs equipped with free potentials $ \Phi $ too. Then the ISDEs become  
\begin{align}\label{:isde2}&
dX_t^i = dB_t^i - \frac{\beta}{2}\nabla \Phi (X_t^i) dt 
- \frac{\beta }{2} \lim_{r\to\infty }
\sum_{j\not=i,\, | X_t^i-X_t^j |<r } 
\nabla \Psi _{\gamma } (X_t^i-X_t^j) dt \quad (i\in\mathbb{N})
.\end{align}

\noindent 
{\bf Definition 1. } 
A solution $ \mathbf{X}=(X_i)_{i\in\mathbb{N}}$ of the ISDE \eqref{:isde2} is called a Coulomb interacting Brownian motion 
if $ d \le \gamma \le d + 2$, and 
a strict Coulomb interacting Brownian motion if $ \gamma = d $. 

\bigskip 

Since $ \gamma \le d+2$, Coulomb interaction potentials are not of Ruelle's class, and one can not apply the classical theory to these potentials. 
The construction of Coulomb interaction Brownian motions 
is a difficult problem. 
Indeed, at present, the only translation invariant strict Coulomb interacting Brownian motion successfully constructed 
is Ginibre interacting Brownian motion; 
there exist no other examples rigorously established. In this case, we have 
\begin{align*}&
(\beta , \gamma , d) = (2,2,2)
.\end{align*}

As for (non-strict) Coulomb interacting Brownian motions, we have 
examples of Coulomb interacting Brownian motions such that 
Dyson's model infinite-dimensions ($ \beta=1,2,4$), Airy interacting Brownian motions ($ \beta = 1,2,4$), and Bessel interacting Brownian motion 
($ \beta=2$) (see \cite{o.rm,o.isde,o.rm2,o-h,o-t.airy,o-t.tail}). 
These except Airy are the solutions of ISDEs \eqref{:isde2} with 
$ d=1 $, $ \gamma = 2 $, and a suitably chosen $ \Phi $. 
The ISDEs in the case of Airy interacting Brownian motions are more complicated than \eqref{:isde2}. We refer to \cite{o-t.airy, o-t.core} for the exact shape of the ISDE. 

To solve these ISDEs, 
we introduced the notions of logarithmic derivative, quasi-Gibbs measures, and the natural coupling among countably many Dirichlet forms describing 
$ k$-labeled processes for all $ k\in\{ 0 \}\cup\mathbb{N} $ 
in \cite{o.isde,o.rm, o.tp}. The resulting solutions were 
(weak) solutions of ISDEs and the uniqueness of the solutions were left open. 

We now find several novel ideas and refine our method to obtain a unique, strong solution for these ISDEs. Namely, we construct strong solutions of ISDEs and prove their strong uniqueness \cite{o-t.tail, o-t.airy}.  

\bigskip 

We next introduce the notion of (resp.\! strict) Coulomb random point fields. 
These random point fields are equilibrium states associated with the unlabeled stochastic dynamics of (resp.\! strict) Coulomb interacting Brownian motions.  

\medskip 

\noindent 
{\bf Definition 2. } 
A random point field $ \mu $ on $ \mathbb{R}^d$ is called 
a Coulomb random point field $ \mu$ if its logarithmic derivative 
$ \mathsf{d}^{\mu} $ is given by ($ \mathsf{s}=\sum_i\delta_{s_i} $)
\begin{align}\label{:22}&
\mathsf{d}^{\mu} (x,\mathsf{s}) = - \beta \{
\nabla \Phi (x) + \lim_{r\to\infty} \sum_{|x-s_i|<r} 
\nabla \Psi _{\gamma }(x-s_i) \} 
\quad \text{ locally in } L^1(\mu^{[1]})
\end{align}
and $ d \le \gamma \le d + 2 $. If in addition $ \gamma = d$, then 
$ \mu $ is called a strict Coulomb random point field. 
Here $ \mu ^{[1]}$ is the 1-Campbell measure of $ \mu $ and 
$ \Phi $ is a free potential and $ \Psi _{\gamma }$ is a Coulomb potential defined by \eqref{:P2}.

\begin{rem}\label{r:22} \thetag{1} See \cite{o.isde} for the definition of the logarithmic derivative of $ \mu $. \\\thetag{2} 
A Coulomb random point field is also called a Coulomb point process. 
\\\thetag{3} 
The convergence in \eqref{:22} is a conditional convergence in general. 
Hence we need $ \lim_{r\to\infty}$ in front of the sum in \eqref{:22}.  
\end{rem}

Though one can no longer use the DLR equation to define Coulomb random point fields, we can still define such random point fields through logarithmic derivatives introduced in \cite{o.isde}. As in the case of 
the stochastic dynamics, 
an only strict Coulomb random point field at present is Ginibre random point field, namely the case $ (\beta,\gamma,d) =(2,2,2)$. 
It is known that Ginibre random point field is a thermodynamic limit of the distributions of the eigenvalues of non-hermitian Gaussian random matrices.

\bigskip

Coulomb interaction potentials have quite strong effect at infinity. 
Hence the feature of associated stochastic dynamics are very different that of the interacting Brownian motions with Ruelle's class potentials. 
As an instance, we present the dynamical rigidity of the Ginibre interacting Brownian motions.

\medskip

The first dynamical rigidity of Ginibre interacting Brownian motion is as follows. 
\begin{thm} [{\cite{o.isde,o-t.tail}}]\label{l:strong} 
Ginibre interacting Brownian motion $ \mathbf{X}$ is a {\em strong} 
solution of the plural ISDEs: 
\begin{align}\label{:strong1}&
dX_t^i = dB_t^i + \lim_{r\to\infty} \sum_{|X_t^i-X_t^j|<r, i\not=j}
\frac{X_t^i-X_t^j}{|X_t^i-X_t^j|^2}
\end{align}
and 
\begin{align}\label{:strong2}&
dX_t^i = dB_t^i - X_t^i 
+ \lim_{r\to\infty} \sum_{|X_t^j|<r, i\not=j}
\frac{X_t^i-X_t^j}{|X_t^i-X_t^j|^2}
.\end{align}
\end{thm}

This result was obtained in \cite{o.isde} at the level of the (weak) solution. 
In \cite{o-t.tail} we refine this result at the level of the unique strong solution. This result means that the {\em real support} of the Ginibre random point field is a very thin set in the configuration space, 
and the configuration 
$ \mathsf{X}_t = \sum_{i} \delta_{X_t^i}$ 
consisting of infinitely many particles $ \{X_t^i\}_{i\in\mathbb{N}}$
 only move this thin set {\em randomly} and {\em rigidly}. 

\bigskip

Let $ \mathsf{S}$ be the configuration space over 
$ \mathbb{R}^2$, and let $ \mu $ be the Ginibre random point field. 
Let $ \ell =(\ell _i)_{i\in\mathbb{N}} $ be a label. 
Let $ \mu_a $ be the reduced Palm measure of $ \mu $ conditioned at $ a $. 
We assume:  
\begin{align*}&
\mu (\, \cdot \, | \ell ^i(\mathsf{s}) = a,\, \mathsf{s}(\{ a \} ) \ge 1) 
\prec 
\mu (\, \cdot \, |  \mathsf{s}(\{ a \} ) \ge 1)) 
\quad \text{ for all }i \in \mathbb{N},\, a\in \mathbb{R}^2 
.\end{align*}
Here $ \mu_1 \prec\mu_2$ means that $ \mu_1$ is absolutely continuous with respect to $ \mu_2$. 

Let $P_{\mathbf{s}} $ denote the distribution of the solution $ \mathbf{X}=(X^i)_{i\in \mathbb{N}}$ of \eqref{:strong1} starting at 
$ \ell (\mathsf{s}) = \mathbf{s}=(s_i)_{i\in\mathbb{N}}$.

\bigskip

We next proceed to the second dynamical rigidity obtained in \cite{o.sub}. 

\begin{thm} [\cite{o.sub}] \label{l:2}
For $ \mu $-a.s.\, $ \mathsf{s}$, and each $ i\in \mathbb{N}$ 
\begin{align}\label{:2}&
\lim_{ \epsilon \to 0}  \epsilon X_{t/ \epsilon ^2}^i = 0 
\quad \text{ weakly in } C([0,\infty);\mathbb{R}^2) 
\text{ $ P_{\ell (\mathsf{s})}$-a.s.}
.\end{align}
\end{thm}

\bigskip

\noindent {\em Remark} 
\thetag{1} In the case of Ruelle's class potentials (with convex cores), 
the limit self-diffusion matrices are always strictly positive definite 
if $ d \ge 2 $ (see \cite{o.p}). 
Hence the result in \tref{l:2} are very different from the result of 
such a standard class. 
\\
\thetag{2} 
The proof of \tref{l:2} is based on the results of geometric rigidity of Ginibre random point field obtained in 
\cite{o.restore-palm, o.sub, o-shirai.palm}. 
\\
\thetag{3} Our argument may be regarded as an infinite-dimensional counter part of the Nash's result of the diagonal estimate of the heat kernel, that deduces the diffusivity/sub-diffusivity of the particles. 
We use various kinds of the geometric rigidity of 
Ginibre random point fields obtained in \cite{o.restore-palm, o-shirai.palm} 
instead of Nash's inequalities in finite dimensions.

\bigskip

We generalize the notion of Coulomb potential  as follows:

Let $ \mathsf{G}_{\gamma}$ be the fundamental solution of $-\frac{1}{2}\Delta $
on $ \mathbb{R}^{\gamma }$. Then by definition 
\begin{align}\label{:a1}&
\mathsf{G}_{\gamma} (x)= 
\begin{cases}
  \frac{2}{\sigma_{\gamma}} 
\frac{1}{\gamma -2} |x|^{2-\gamma }
&(\gamma \not= 2)
\\
-\frac{2}{\sigma_{\gamma}}\log |x|
& (\gamma = 2)
.\end{cases}
\end{align}
Here $ \sigma_{\gamma} = 2\pi^{\gamma /2}/\Gamma (\frac{\gamma }{2})$ 
 is the surface volume of the $ (\gamma -1) $-dimensional surface: 
$ \{x\in\mathbb{R}^{\gamma } ;|x|=1 \} $. Its gradient is then given by 
\begin{align}\label{:a2}&
\nabla \mathsf{G}_{\gamma} (x)= 
 - \frac{2}{\sigma_{\gamma}} \frac{x}{|x|^{\gamma }} 
.\end{align}
Note that $ \sigma_{1}=2$ and $ \sigma_{2}=2\pi $. Hence we deduce that 
\begin{align}& \label{:P1}
\nabla \mathsf{G}_{1} (x)= - \frac{x}{|x|} \quad \text{ and } \quad 
\nabla \mathsf{G}_{2} (x)= - \frac{1}{\pi } \frac{x}{|x|^2}
.\end{align}
We now set 
\begin{align}\label{:P2}&
\Psi _{\gamma} (x) = 
\frac{\sigma_{\gamma}}{2} \mathsf{G} _{\gamma }(x)
.\end{align}
Then we see by definition that 
\begin{align}& \label{:P3}
\nabla \Psi _{\gamma} (x) =  - \frac{x \, }{|x|^{\gamma }} 
.\end{align}

Since $ \Psi _{\gamma }$ gives an electrostatic potential 
in $ \mathbb{R}^{\gamma }$ ($ \gamma = 3$),  we call it a Coulomb potential. 
The sign of $ \Psi _{\gamma }$ is chosen in such a way that the potential describes the system of one component plasma.

We thus call a random point field 
$ \mu _{\beta,\gamma,d}$ in $ \mathbb{R}^d$ 
 a Coulomb random point field if its logarithmic derivative 
 $ \mathsf{d}^{\mu}$ is given by 
 \eqref{:22} with $ \gamma $-dimensional Coulomb potential $ \Psi _{\gamma }$ such that $ d \le \gamma \le d + 2 $ 
and inverse temperature $ \beta > 0 $. 
We call $ \mu _{\beta,\gamma,d}$ 
a strict Coulomb random point field if $ \gamma = d $ in addition.

We remark again that 
Ginibre random point field is an only example of translation invariant, 
 strict Coulomb random point fields rigorously constructed, 
and is the case of $ (\beta,\gamma,d) = (2,2,2)$.

We finally note that one can generalize $ \gamma \in \mathbb{N}$ to any positive numbers and define $ \Psi _{\gamma }$ by $ \nabla \Psi _{\gamma } 
 = - x/|x|^{\gamma } $.
If $ d+2 < \gamma $, then $ \Psi _{\gamma }$ is a potential in the regime to which the classical theory can be applied. 
 In the case of $ d \le \gamma \le d + 2$, this is not the case, and 
$ \Psi _{\gamma }$ is interesting enough to study.

{\small

\noindent 
Acknowledgement: 
H.O. is supported in part by the Grant-in-Aid for Scientific Research (KIBAN-A, No. 24244010) and the Grant-in-Aid for Scientific Research (KIBAN-B, No. 21340031)@

\end{document}